# Formation and Thermodynamic Evolution of plasmoids in active region jets

Sargam M. Mulay,[1]⋆ Durgesh Tripathi,[2]† Helen Mason,[3]‡ Giulio Del Zanna,[3]§ & Vasilis Archontis[4,5]¶

[1] *School of Physics & Astronomy, University of Glasgow, G12 8QQ, Glasgow, UK*
[2] *Inter-University Centre for Astronomy and Astrophysics, Post Bag-4, Ganeshkhind, Pune 411007, India*
[3] *DAMTP, Centre for Mathematical Sciences, University of Cambridge, Wilberforce Road, Cambridge, CB3 0WA, UK*
[4] *School of Mathematics and Statistics, University of St Andrews, North Haugh, St Andrews, Fife KY16 9SS, UK*
[5] *Section of Astrogeophysics, Department of Physics, Laboratory of Astronomy, University of Ioannina, 45110 Ioannina, Greece.*

**ABSTRACT**

We have carried out a comprehensive study of the temperature structure of plasmoids, which successively occurred in recurrent active region jets. The multithermal plasmoids were seen to be travelling along the multi-threaded spire as well as at the footpoint region in the EUV/UV images recorded by the Atmospheric Imaging Assembly (AIA). The Differential Emission Measure (DEM) analysis was performed using EUV AIA images, and the high-temperature part of the DEM was constrained by combining X-ray images from the X-ray telescope (XRT/Hinode). We observed a systematic rise and fall in brightness, electron number densities and the peak temperatures of the spire plasmoid during its propagation along the jet. The plasmoids at the footpoint (FPs) (1.0–2.5 MK) and plasmoids at the spire (SPs) (1.0–2.24 MK) were found to have similar peak temperatures, whereas the FPs have higher DEM weighted temperatures (2.2–5.7 MK) than the SPs (1.3–3.0 MK). A lower limit to the electron number densities of plasmoids - SPs (FPs) were obtained that ranged between 3.4–6.1×$10^8$ (3.3–5.9×$10^8$) cm$^{-3}$ whereas for the spire, it ranged from 2.6–3.2×$10^8$ cm$^{-3}$. Our analysis shows that the emission of these plasmoids starts close to the base of the jet(s), where we believe that a strong current interface is formed. This suggests that the blobs are plasmoids induced by a tearing-mode instability.

**Key words:** Sun: atmosphere – Sun: activity – Sun: corona – Sun: evolution – Sun: UV radiation – Sun: X-rays

## 1 INTRODUCTION

Many small-scale events like solar jets (e.g., Chifor et al. 2008; Raouafi et al. 2016; Shen 2021; Mulay et al. 2016), UV bursts (e.g., Peter et al. 2014; Gupta & Tripathi 2015; Rouppe van der Voort et al. 2017) as well as large-scale energetic, eruptive events like solar flares (e.g., Milligan et al. 2010), coronal mass ejections (CMEs) (e.g., Ko et al. 2003; Mulay et al. 2014; Patel et al. 2020; Bemporad et al. 2022), erupting flux ropes (e.g., Liu 2013; Tripathi et al. 2006, 2007, 2009, 2013) in the solar atmosphere are thought to be produced by magnetic reconnection (Parker 1963; Petschek 1964). Traveling dense plasma structures known as magnetic islands or plasmoids have been observed in many features, including hot loops associated with flares (Takasao et al. 2012), thin current sheets formed after a CME (Lin et al. 2005; Guo et al. 2013), and along solar jets (Alexander & Fletcher 1999; Zhang & Ji 2014b; Lu et al. 2019). It is suggested that plasmoids can travel upward and downward, become part of the nearby plasma structures and/or disappear.

Plasmoids are thought to be produced by a tearing mode instability at the current sheet region (Moreno-Insertis & Galsgaard 2013; Yang et al. 2013; Wyper et al. 2016). Kumar et al. (2019) reported the first evidence of plasmoid formation during breakout reconnection in two successive coronal jets in the fan-spine topology of an embedded bipole. Yokoyama & Shibata (1994, 1996) reported the existence of cool, dense magnetic islands in the 2D numerical simulations of solar coronal X-ray jets by solving the resistive magnetohydrodynamic (MHD) equations. Leake et al. (2012) successfully produced these chromospheric plasmoids in the simulation using a partially ionized reacting multi-fluid plasma model. The authors reported that a secondary tearing instability known as plasmoid instability is responsible for plasmoid formation.

Plasmoids have also been observed along recurrent jets, suggesting that repetitive reconnection occurs at their footpoints (Alexander & Fletcher 1999; Zhang & Ji 2014a; Mulay 2018). They were observed at low chromospheric temperatures (Singh et al. 2011, 2012), at transition region temperatures (Mulay et al. 2017a; Zhang & Ni 2019; Joshi et al. 2020) and at coronal temperatures using EUV (Alexander & Fletcher 1999; Zhang & Ji 2014b; Chen et al. 2022) and X-ray (Mulay 2018) observations. Singh et al. (2012) studied chromospheric anemone jets using the Ca II H filtergram of the Solar Optical Telescope (SOT) onboard Hinode. They observed recurrent ejections of bright plasmoids along the jet, which were traveling downward with a speed of 35 km s$^{-1}$ and interpreted them as a result of chromospheric magnetic reconnection and the tearing mode instability.

⋆ E-mail: Sargam.Mulay@glasgow.ac.uk
† E-mail: durgesh@iucaa.in
‡ E-mail: hm11@damtp.cam.ac.uk
§ E-mail: gd232@damtp.cam.ac.uk
¶ E-mail: va11@st-andrews.ac.uk

Mulay et al. (2017b) reported UV observations of bright and compact plasmoids along recurrent jets using Si IV 1400 Å images obtained from the Slit-Jaw imager (SJI) on board the Interface Region Imaging Spectrograph (IRIS). The plasmoids were seen to be following the helical motion of the jet spire. Their coronal counterparts were observed in the 171 Å channel of the Atmospheric Imaging Assembly (AIA; Lemen et al. 2012) instrument on board the Solar Dynamics Observatory (SDO; Pesnell et al. 2012). Similar to Mulay et al. (2017b), Zhang & Zhang (2017); Zhang & Ni (2019) and Joshi et al. (2020) also reported observations of plasmoids with sizes of 0.45″–1.35″, travelling with variable speeds (10 to more than 220 km s$^{-1}$) along the UV jets as seen in C II 1330 Å SJIs.

Similar plasmoids were previously identified in EUV jets by Alexander & Fletcher (1999) using the Transition Region and Coronal Explorer (TRACE; Handy et al. 1999) observations with a timescale of 2-3 min. AIA observations of plasmoids in jets were also reported by Zhang & Ji (2014b) and Zhang et al. (2016b). Zhang & Ji (2014a) measured the size of the plasmoids to be ~3 Mm in diameter with lifetimes of 24-60 s whereas Zhang et al. (2016a) found bigger plasmoids of sizes 4.5–9 Mm traveling at 140–380 km s$^{-1}$. Filippov et al. (2015) found smaller average speeds for plasmoids along the EUV jet, about 80 km s$^{-1}$. Chen et al. (2022) studied five plasmoids along the jet and found their widths to range between 0.8 and 2.3 Mm, with apparent velocities from 59 km s$^{-1}$ to 185 km s$^{-1}$.

In spite of having their signatures observed at various wavelengths/temperatures, a complete temperature structure of the plasmoid is still unknown. This is partly due to the limited temperature coverage of EUV/UV broad-band imagers, and partly due to a lack of accurate atomic modeling to calculate the temperature responses. A few individual studies (Zhang & Ji 2014b; Zhang et al. 2016b; Mulay 2018; Chen et al. 2022) attempted to determine the temperature of the plasmoids using broadband EUV images and a differential emission measure (DEM) analysis. Zhang & Ji (2014b) and Zhang et al. (2016b) measured a similar temperature of plasmoids, ranging between 1.8–3.3 MK, with densities 1.7–3.3×10$^9$ cm$^{-3}$. Chen et al. (2022) reported an effective temperature of the plasmoids to be 24 MK.

Mulay (2018) combined EUV images from AIA with X-ray data from the X-ray Telescope (XRT; Golub et al. 2007) for a DEM analysis of plasmoids along jets and found peak temperatures to be in the range 1.6–2.5 MK, while the DEM-weighted averaged temperatures were found to be 2.5–4 MK. These temperatures are similar to those obtained for the jet spire region by Mulay et al. (2017a). In addition, they obtained a lower limit on the electron number densities $N_e$, which ranged from 3.1 to 4.8 ×10$^8$ cm$^{-3}$.

In this paper, we extend the Mulay (2018) study by improving the temperature analysis and focusing on the thermodynamical properties of the plasmoids along jets and their evolution. For the DEM analysis, we use more confined boundaries around the plasmoid which helped to reduce the background emission. Note that these are not background-subtracted images. The background included is minimal compared to the boxed region defined around plasmoids in Mulay (2018). In addition, we derive the temperatures for the plasmoids that were observed at the footpoints of the jets, and we compare their temperatures with those of the spire. The evolution of the temperature of one of the spire-plasmoid is studied thoroughly by tracking its movement along the spire.

The rest of the paper is structured as follows: UV/EUV and X-ray imaging observations are discussed in §2. The thermodynamic structure of the plasmoids and physical parameters are discussed in §3. We finally summarize and conclude in §4.

## 2 OBSERVATIONS

The active region NOAA #11330 produced homologous and recurrent jets from its western periphery between 13:00 and 18:00 UT (see AIA 171 Å movie online in the supplementary material of Mulay et al. 2017a). A blob-like (hereafter referred to as plasmoid) structure was first noticed at 13:49 UT. After that, a continuous stream of plasmoids along the jet spire was observed between 14:30 and 15:30 UT. In this work, we focus on this one-hour jet activity and study the behaviour of these plasmoids.

We have used data recorded between 14:30 to 15:30 UT by AIA in its six EUV channels (94 Å, 171 Å, 131 Å, 193 Å, 211 Å, 335 Å), one UV (1600 Å) channel and XRT data in the Ti-poly and Be-thin filters. Both the UV and EUV AIA images have a pixel size of ~ 0.6″, but the temporal resolution is 12 s for the EUV and 24 s for the UV channels. The XRT images have a pixel size of 1.02″. The temporal resolution of the Ti-poly images is 1 min, whereas that for the Be-thin filter is 15 min.

For our analysis, we downloaded full disc level 1 AIA data from the Joint Science Operations Center (JSOC[1]). The XRT level 1 data were obtained from the Hinode Science Data Centre Europe[2]. The initial data processing was carried out using the standard routines available in the SolarSoftWare (SSW; Freeland & Handy 1998) libraries. The intensities were normalized using the exposure times. We used the XRT calibration by Narukage et al. (2011). The AIA and XRT images were co-aligned using the procedure given in Yoshimura & McKenzie (2015).

Figure 1 (left panel) displays a composite image of the active region and the associated jet. The composite image is created using co-temporal images taken in AIA 171 (red), 193 (green), and 211 Å (blue) channels. The AIA images in these channels are processed using the multiscale Gaussian normalization method by Morgan & Druckmüller (2014) to bring out fine-scale features within the active region, and then we created a composite image. The over-plotted yellow box locates the observed jet. The jet spire and footpoint are labeled. Note the curved structure of the spire (near-parabolic trajectory) and inverted Y-shape at the footpoint. A zoomed-in view of the jet, corresponding to the field-of-view (FOV) covered by the yellow box in the left panel, is shown in the right panel. The bright plasmoids along the spire and the footpoints of the jet are indicated by white and green arrows, respectively. This is a unique observation that shows plasmoids along the spire as well as at the footpoints of the jets.

Figure 2 (see online Movie 1) shows near-simultaneous AIA and XRT images of the jet, as shown in the right panel of Fig 1. The inverted Y-shape of the footpoint of the jet with a curved spire is conspicuously observed in all EUV channels. In the image taken by AIA in 1600 Å and the XRT images, only the base of the jet is observed, which appeared slightly brighter in the AIA 1600 Å. Considering that the base and the lower part of the jet spire are observed in 1600 Å and in all the six EUV channels of AIA, as well as in both XRT filters, it is clear that the emission is multi-thermal, similar to the observations of Mulay et al. (2016); Tripathi (2021).

Figure 2 further reveals that the bright plasmoids with well-defined boundaries along the spire are clearly observed in the AIA 335, 193, 211, and 171 Å channels. However, in 94 and 131 Å channels, the plasmoids did not appear clearly at this timing, albeit they could be identified by a closer look at the variation in the intensities. These plasmoids appear clearer at later times in the AIA 94 and 131 Å images. The existence of similar morphology is likely to indicate that they are the same plasmoids that appeared earlier in other channels. We find that the jet spire is much fainter and thinner in the 1600 Å and XRT images than in other AIA EUV channels. Note that 1600 Å and XRT images displayed in Fig. 2 is taken ∼10 s earlier than those in the EUV. A multi-thermal nature of the spire, footpoints, and plasmoids is evident from Figs. 1 and 2.

---

[1] http://jsoc.stanford.edu/
[2] http://sdc.uio.no/sdc/

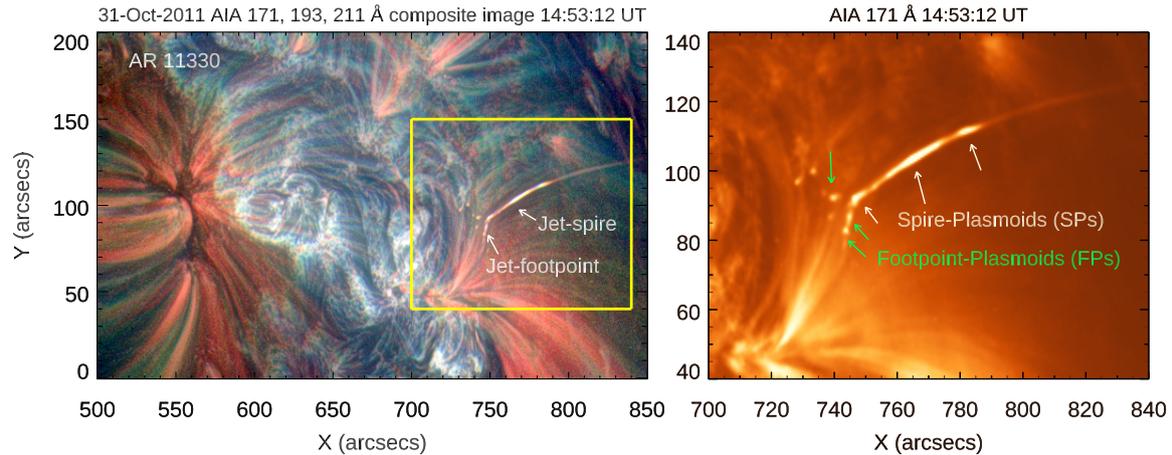

**Figure 1.** Left panel: A composite image of the active region NOAA #11330, and the AR jet. The jet spire and jet footpoint are labelled. The image is created using three AIA 171 Å (red), 193 Å (green), and 211 Å (blue) channels. The yellow box shows the region of interest displayed in the right panel. Right panel: A zoomed-in view of the region of interest boxed in the left panel in the AIA 171 Å channel. The plasmoids along the jet spire (SPs) and footpoints (FPs) are indicated by white and green arrows, respectively.

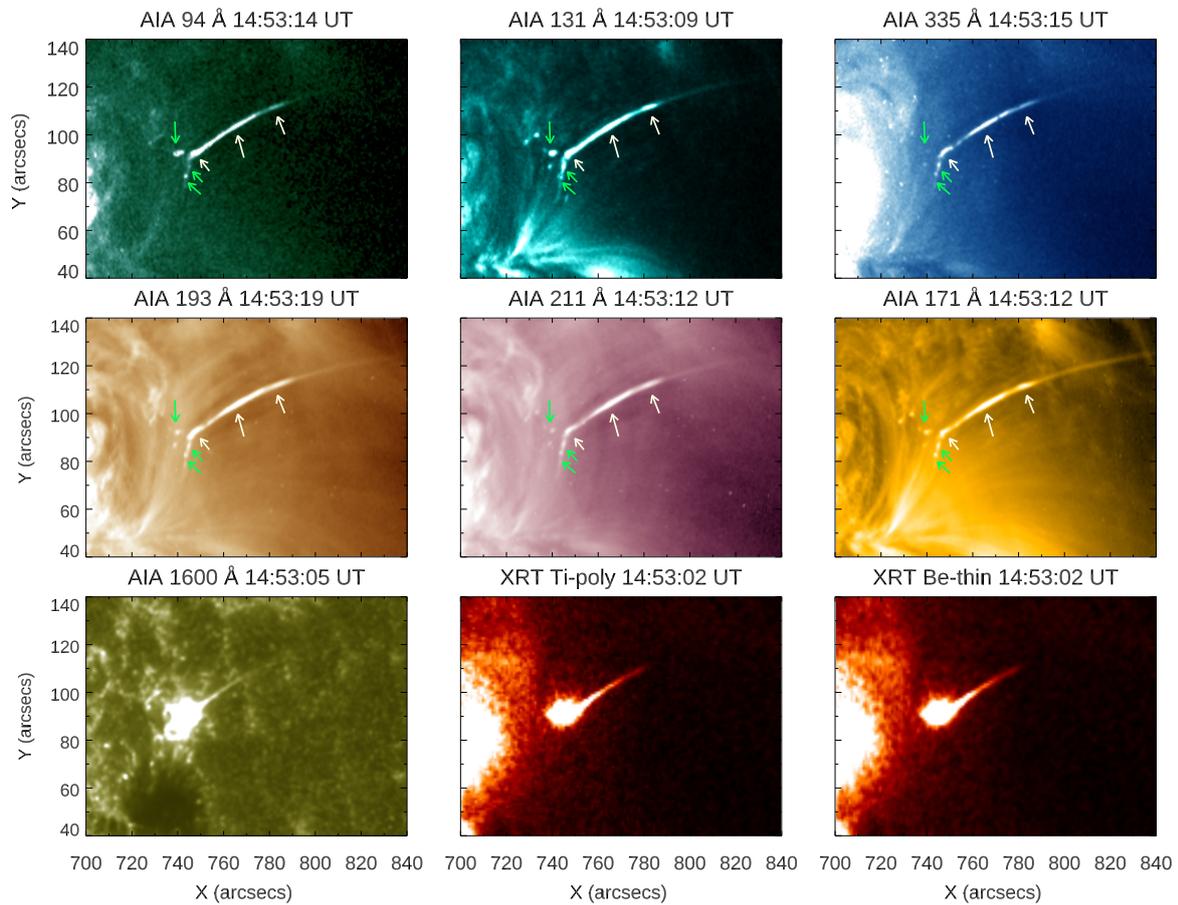

**Figure 2.** Near simultaneous observation of the jet shown in the right panel of Fig. 1 in AIA and XRT. The plasmoids along the jet spire and footpoints are indicated by the white and green arrows, respectively. See the evolution of the AR jet in Movie 1.

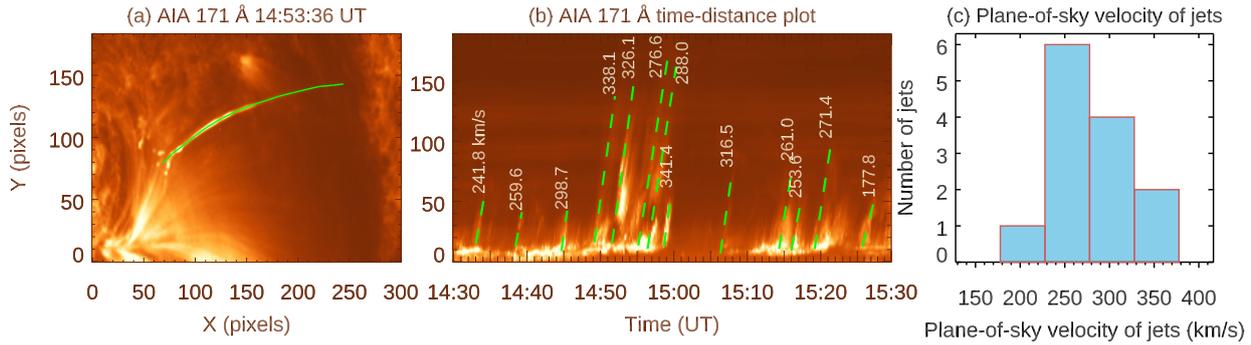

**Figure 3.** Panel (a): One of the recurrent jets was observed at 14:53:36 UT in the AIA 171 Å channel. The green solid line along the spire of the jet indicates the position of a curved slit. Panel (b): a time-distance plot for homologous and recurrent jets for the period of 1 hour from 14:30 to 15:30 UT. The plane-of-sky velocities for the jets were measured along the dashed green lines, and the numbers indicate velocities in km s$^{-1}$. Panel (c): Histogram of the plane-of-sky velocities of the jets.

### 2.1 Evolution of plasmoids along the jet spire

The first bright jet with a curved spire appeared in all AIA and XRT channels at 14:30:02 UT (see the evolution of AR jet in Movie 1). The spire of the jet reached a certain height and started to disappear at 14:31:21 UT. Before its complete disappearance, another jet from the same footpoint region was produced at 14:32:26 UT, with a slightly wider spire than the previous jet. The second jet showed multiple threads, in addition to a small plasmoid at 14:34:24 UT, which disappeared completely in 24 s. Note that there was no jet activity observed for 4 min (i.e. until 14:38:36 UT). Moreover, the intensity of the bright footpoint slowly decreased during this time. The footpoint started to become brighter at 14:38:24 UT and produced a series of thin jets.

A series of images taken between 14:30 and 15:30 UT show continuous episodes of appearance and disappearance of plasmoids. These images show that most of the plasmoids appear very close to the footpoint of the jet and travel along the spire. Most of them travel for short distances before they either disappear or merge into the spire of the jet.

## 3 DATA ANALYSIS AND RESULTS

### 3.1 Speed of jets

We measured the plane-of-sky velocities for all recurrent jets observed during 14:30-15:30 UT using time-distance maps of the AIA 171 Å channel. All jets were homologous and exhibited curved spires. Therefore, we selected an artificial curved slit along the spire having a width of three AIA pixels. The jet-front was tracked and the intensity along this slit was averaged for each AIA image and plotted as a function of time.

We display the observations of the jet taken at 14:53:56 UT in AIA 171 Å in panel (a) of Fig. 3. The over-plotted solid green line indicates the position of the curved slit along the jet spire. Panel (b) shows the time-distance plot obtained using the curved slit. The slanting bright stripes are the recurrent jets. Note that it was difficult to identify individual plasmoids along the spire as well as along the footpoint in the time-distance map due to the averaging of the intensities along the three pixels wide slit. The curved slit captured all emission along the spire. The green dashed lines along different jets are used to measure the plane-of-sky velocities, which range between 178–341 km s$^{-1}$. These velocities are similar to those measured by Shibata et al. (1994), Schmieder et al. (2013), Chandra et al. (2015), Mulay et al. (2016), Solanki et al. (2020) but are larger than those reported by Wang et al. (2018) and Mulay et al. (2019). They are smaller than those reported by Mulay et al. (2017a) and Mulay (2018), which could be attributed to the fact that a straight slit was used along the curved spire. This might not have captured all the emission along the curved spire and resulted in higher velocities. Fig. 3 in panel (c) shows a histogram of the plane-of-sky velocities shown in panel (b). It shows that most jets have plane-of-sky velocities between 225 and 275 km s$^{-1}$.

### 3.2 Temperatures of the spire and footpoint plasmoids

In order to deduce the temperature structure of the plasmoids, we used the Differential Emission Measure (DEM) inversion method developed by Weber et al. (2004). We obtained DEM estimates between log $T$ [K] = 5.5 and 7.2 (0.3-15.8 MK). To better constrain the high temperatures, we used near-simultaneous XRT Ti-poly images along with the six EUV images from AIA. As the AIA and XRT images have different pixel sizes, we have re-binned the AIA images to the XRT resolution (see Fig. 4 and 7).

The response functions of AIA and XRT were computed using the CHIANTI atomic database (Dere et al. 1997, 2019) following the procedure outlined by Del Zanna et al. (2011). Moreover, we used the photospheric abundances by Asplund et al. (2009), and the ionization equilibrium calculations by Dere et al. (2009). Mulay et al. (2017a) studied the same active region jet event using Hinode/EIS observations. The authors used the intensity of two Fe XII lines (at log $T$

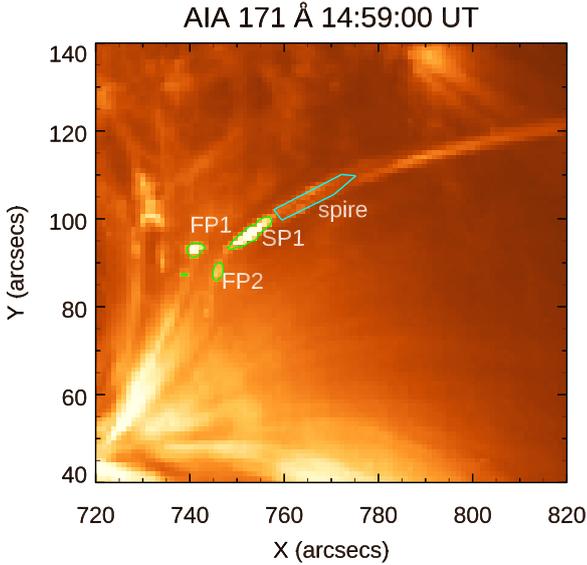

**Figure 4.** AIA 171 Å image (a re-binned AIA image to XRT resolution) of the jet at 14:59:00 UT. The plasmoids at the footpoints (FP1, FP2) and along the jet spire (SP1) are indicated by green contours. The cyan contour indicates emission from the spire region

[K] = 6.2 i.e. 1.6 MK) ($\lambda$(186.854+186.887)/$\lambda$(195.119+195.179)) from Hinode/EIS and the density diagnostic technique to measure the electron number density at the spire and footpoint region of the jet. They obtained the electron number density of $7.6 \times 10^{10}$ cm$^{-3}$ for the spire region from the density diagnostic and we used this value to calculate the response function of AIA channels.

To identify the plasmoids, we draw iso-intensity contours on the 171 Å images. Fig. 4 displays the jet observed at 14:59:00 UT. The over-plotted iso-intensity contours show FP1, FP2, and SP1, where FP and SP stand for "footpoint-plasmoid" and "spire-plasmoid", respectively. The intensities in the plasmoids were then obtained by averaging over the areas covered by these contours. We used these regions to obtain intensities in the other AIA and XRT channels. This selection of pixels improved the DEM analysis by reducing the background emission that was present in the larger regions that were selected in Mulay (2018) for the analysis of the same plasmoids. In addition to the plasmoids, we have also identified a region along the spire (shown as cyan contour and labelled as a spire in Fig. 4) for comparison.

To study the temperature evolution, we selected six timings (see Table 2) when the plasmoids and spire were clearly observed. We performed a DEM analysis using near-simultaneous AIA and XRT images during these six periods but discussed in detail the plasmoids observed at 14:59 UT and shown in Fig. 4, namely FP1, FP2, and SP1. Note that SP1 is brighter than both FP1 and FP2 and the spire region (See Table 1). Fig. 5 displays the DEM plots for FP1 (panel (a)), FP2 (panel (b)), SP1 (panel (c)) and the spire (panel (d)). The best-fit DEMs are shown in black. To assess uncertainties in the DEMs, we have obtained several Monte-Carlo solutions by varying the input intensities. The colour bars represent the number of solutions (yellow bars 50%, brown bars 80%, and blue bars 95%) in each temperature bin. The predicted intensities for all regions were found to be in good agreement with the observed intensities, as shown in Table 1. Large uncertainties in the DEMs are present for temperatures below 0.5 MK and above 5 MK, as the observations do not provide enough constraints. The parameters obtained from the DEMs are listed in Table 2, panel (f).

Figure 5 reveals that for FP1, FP2, SP1 and spire regions, the DEM peaks at log $T$ [$K$] = 6.1 (1.3 MK), 6.30 (2.0 MK), 6.35 (2.2 MK) and 6.25 (1.8 MK), respectively. The footpoint plasmoids have lower temperatures compared to the spire-plasmoid. The DEM for FP1 (panel a) towards higher temperatures is not well constrained. The DEM for FP2 (panel b) falls slowly at low-temperatures, log $T$ [$K$] = 5.8 (0.6 MK) but falls rapidly at high-temperatures, log $T$ [$K$] = 6.6 (4.0 MK). The SP1 DEM (panel c) has a peak at log $T$ [$K$] = 6.35 (2.2 MK). The low-temperature part of the DEM is not well constrained and shows high DEM values, whereas the high-temperature part of the DEM does not fall as sharply as the FP2 DEM. The spire DEM (panel d) show a peak at log $T$ [$K$] = 6.25–6.35 (1.8–2.2 MK). The high-temperature part of the DEM falls more sharply than the other DEMs.

### 3.2.1 Temperature of plasmoids

We use these DEM estimates to compute the DEM-weighted average temperature ($\tilde{T}$) using

$$\tilde{T} = \frac{\int DEM(T) \, log \, T \, dT}{\int DEM(T) \, dT} \quad (1)$$

Note that for this purpose, we used the DEM values corresponding to the temperature range log $T$ [$K$] = 5.5 and 7.2 (0.3 - 15.8 MK). The DEM-weighted average temperature was found to be log $T$ [$K$] = 6.76 (5.7 MK), 6.36 (2.3 MK), 6.30 (2.0 MK), and 6.34 (2.2 MK) for the FP1, FP2, SP1, and spire respectively. For most of these plasmoids, these DEM weighted averaged temperatures are slightly higher than the peak temperatures in the DEMs.

### 3.2.2 Electron number density of plasmoids

By assuming a filling factor ($f$) equal to unity, we calculated a lower limit to the electron number density ($N_e$) for the plasmoids using the following equation

$$N_e = \sqrt{\frac{EM}{0.83 \, f \, dh}} \quad (2)$$

where $dh$ is the column depth which is taken the same as the width of the plasmoids and spire observed in the AIA 171 Å channel (about 2″). The details of the measurement of plasmoid width are given in the Appendix. The column emission measure (EM) was obtained by integrating the DEM values over the entire temperature range. The obtained densities range between 2.6–5.8×10$^8$ cm$^{-3}$. We note that the electron number density values obtained using DEM measurements represent the lower limit to the electron number density.

We have performed the same analysis on the plasmoids observed in the other five instances. The results obtained are summarized in Table 2 (panels a–e) and in Fig. 6. Taking into account all six cases, we find that the temperatures at DEM peaks for the footpoint plasmoids range between log $T$ [$K$] = 6.0-6.4 (1.0–2.5 MK), whereas the DEM weighted average temperatures ranged between log $T$ [$K$] = 6.34–6.76 (2.2–5.7 MK). For almost all but one footpoint plasmoids, we find that the DEM weighted temperature is higher than the temperature at which the highest DEM is observed. Similarly, the temperature at which the DEM of the spire plasmoids peaks ranges between log $T$ [$K$] = 6.0-6.35 (1–2.24 MK), whereas the DEM weighted average temperatures range between log $T$ [$K$] =

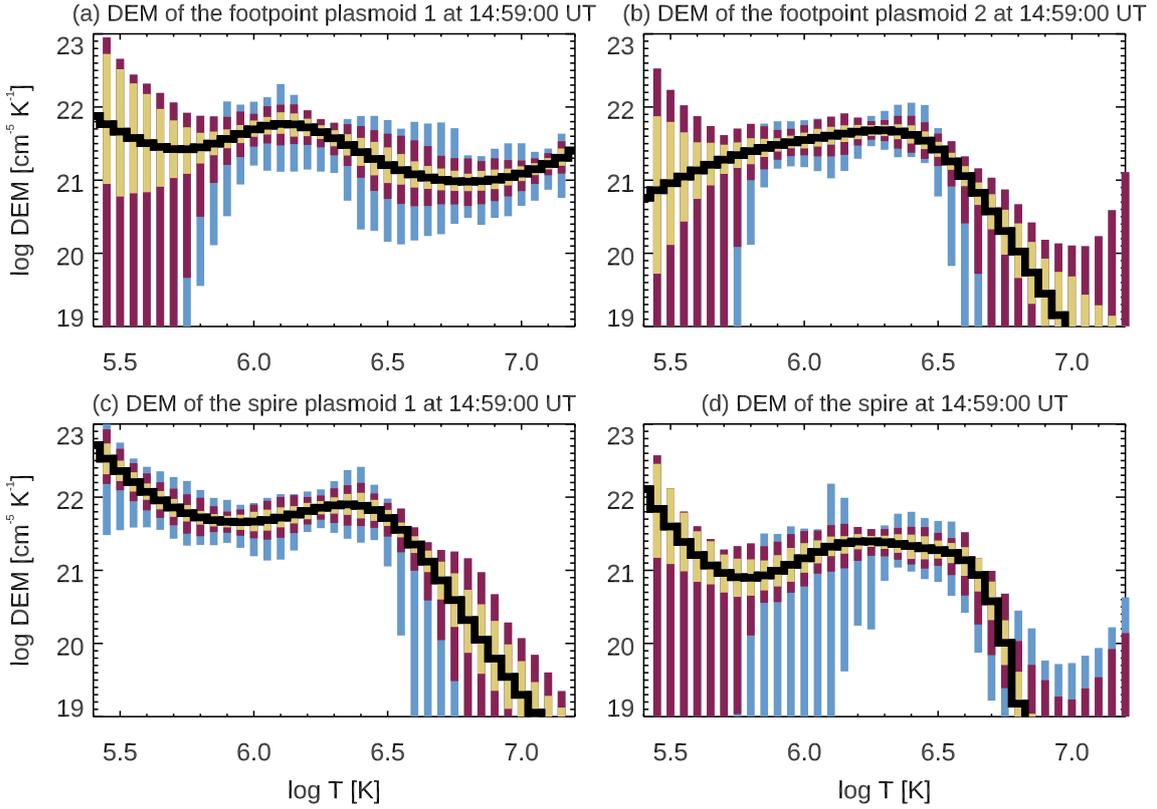

**Figure 5.** DEM plots for the (a) FP1, (b) FP2, (c) SP1 and (d) spire. The best-fit DEMs are shown by black curves. The yellow bars represent 50%, brown bars 80%, and blue bars 95% of the Monte-Carlo solutions in each temperature bin.

**Table 1.** Observed and predicted count rates (DN/s) for the plasmoids observed along the spire and at the footpoint at 14:59:00 UT.

| Column 1 | Column 2 | Column 3 | Column 4 | Column 5 | Column 6 | Column 7 | Column 8 | Column 9 |
|---|---|---|---|---|---|---|---|---|
| AIA/XRT | Footpoint plasmoid 1 | | Footpoint plasmoid 2 | | Spire plasmoid 1 | | Spire | |
| Channel | Observed | predicted | Observed | predicted | Observed | predicted | Observed | predicted |
| (Å) | (DN/s) | (DN/s) | (DN/s) | (DN/s) | (DN/s) | (DN/s) | (DN/s) | (DN/s) |
| AIA 94 | 21.6 | 21.3 (1.01) | 8.0 | 6.7 (1.2) | 28.5 | 11.7 (2.4) | 4.0 | 4.0 (1.0) |
| AIA 131 | 178.8 | 166.4 (1.07) | 61.0 | 67.0 (0.9) | 176.1 | 142.6 (1.2) | 33.2 | 32.6 (1.0) |
| AIA 171 | 1639.5 | 1555.2 (1.05) | 1182.4 | 1232.1 (0.95) | 1948.5 | 2511.09 (0.8) | 596.2 | 595.4 (1.0) |
| AIA 193 | 1473.6 | 1593.2 (0.93) | 1086.1 | 1185.0 (0.91) | 2295.3 | 2967.6 (0.8) | 754.5 | 902.966 (0.8) |
| AIA 211 | 465.6 | 471.6 (0.98) | 433.1 | 439.4 (0.98) | 800.6 | 911.4 (0.9) | 226.1 | 283.3 (0.8) |
| AIA 335 | 25.3 | 25.0 (1.01) | 25.0 | 23.4 (1.1) | 49.8 | 44.7 (1.1) | 16.0 | 14.3 (1.1) |
| Ti-poly | 798.1 | 727.6 (1.1) | 116.0 | 126.0 (0.9) | 200.2 | 234.8 (0.85) | 90.0 | 90.3 (1.0) |

**Notes -** The numbers in parentheses given in Columns 3, 5, 7, and 9 indicate the ratio of observed and predicted intensities in each channel.

6.14-6.47 (1.3–3.0 MK). Similarly to the FPs, we find that the DEM-weighted average temperatures for the SPs are always higher than those for the peak of the DEM. For comparison, we have also derived the DEM and the DEM-weighted average temperatures for the spire. We find that the temperatures of the spire regions were in the range log $T$ [K] = 6.25–6.35 (1.8–2.2 MK) whereas the DEM-weighted average temperatures were found to be between log $T$ [K] = 6.34-6.7 (2.2–5 MK), higher than the temperatures at peak DEMs, as seen for the FPs and SPs. We do not see significant differences between the temperatures of the spires compared to those from FPs and SPs.

The lower limit to the electron number densities of FPs (SPs) ranges between $3.3$–$5.9 \times 10^8$ cm$^{-3}$ ($3.4$–$6.1 \times 10^8$ cm$^{-3}$). For the spire regions, we find densities in the $2.6$–$3.2 \times 10^8$ cm$^{-3}$ range. The densities of the plasmoids are higher than those in the spire of the jet, which explains why they appear brighter.

We studied the same jet using EIS/Hinode observations in Mulay et al. (2017a) and showed the difference in DEM and EM results using the photospheric (Asplund et al. 2009) and coronal abundances (Feldman 1992). Using Skylab observations Widing et al. (1986); Feldman & Laming (2000) pointed out the photospheric composition for the surge event on the limb. In the work by Lee et al. (2015), they reported photospheric abundances for polar jets (in connection with fast solar wind). It is not yet clear whether photospheric abundances would also apply to active region jets. This is a topic of current research and interest with regard to the formation of the slow solar wind. Since we do not have direct measurements of the abundances for active region jets, if we consider coronal abundances by Feldman (1992) in the DEM analysis of plasmoids, the EM, could vary by a factor of 3-4. Hence the electron number density by a factor of two.

### 3.3 Temporal evolution of temperature and density in the plasmoids

Figure 7 displays the temporal evolution of a plasmoid in a sequence of AIA 171 Å images. The plasmoid first appeared at the base of the spire at 14:58:36 UT and then moved along the spire. Initially, the plasmoid was quite faint, but its intensity started to increase as it travelled along the spire. After about 90 s, it showed a decrease in its intensity before it finally disappeared. We also noticed that the size of the plasmoid along the spire varied during its lifetime, as shown in Fig. 7.

We performed a similar DEM analysis as discussed in section 3.2 at the time steps shown in Fig. 7. The DEM plots are shown in Fig. 8 and the colour bars are the same as those plotted in Fig. 5. The time at which these DEM curves were derived is also labelled. Note that in this case, the DEM was only performed in the temperature interval $\log T \, [K] = 5.4 – 6.8$ (0.25–6.3 MK). The temperature was reduced to get a better estimate on the predicted intensities and best-fit DEM.

Except for three instances of time, 14:58:48, 14:59:00, and 14:59:12 UT, the plasmoid showed double-peaked DEMs, one at low and another at high temperatures. Large uncertainties in the DEM values were seen for the temperatures below $\log T \, [K] = 5.7$ (0.5 MK) and above $\log T \, [K] = 6.6$ (5 MK). The observed and predicted intensities in each filter at each instant of time are given in the Appendix in Tables B1 and B2, showing excellent agreement. We have used these DEMs to derive the plasma parameters at each instant of time. The results are given in Table 3. Fig. 9 shows the distribution of plasma parameters obtained for the spire-plasmoid as it travels along the spire. The peak temperatures were found to be initially increasing and then decreasing during its propagation along the spire, whereas the DEM weighted average temperatures do not show a similar variation. The peak temperatures ranged from 1.2-2.24 MK, whereas the DEM-weighted temperatures were slightly higher, 1.6-2.5 MK.

Unlike the temperatures, the densities show a systematic rise and fall as the plasmoid travels, in the range $2.3\text{-}5.0 \times 10^8$ cm$^{-3}$. The density at 14:58:48 UT was $3.6 \times 10^8$ cm$^{-3}$. It increased for every 12 sec time step, reaching $5.0 \times 10^8$ cm$^{-3}$ at 14:59:12 UT, to then decrease to $2.3 \times 10^8$ cm$^{-3}$ at 15:00:12 UT before it disappeared.

## 4 DISCUSSION AND SUMMARY

In this paper, the thermodynamical structure of the plasmoids along the spire and footpoint of a recurrent jet were studied in detail using EUV and X-ray imaging observations from AIA/SDO and XRT/Hinode. The high-cadence observations enabled us to obtain DEMs, peak temperatures, a lower limit to the electron number densities, and DEM-weighted average temperatures of the plasmoids. In summary,

- We observed the multithreaded structure of the arc-shaped spire of the jet. We spotted multithermal plasmoids along the spire and at the footpoint of the jet (See Fig. 2).

- The temperatures of the DEM peaks for the footpoint plasmoids range between $\log T \, [K] = 6.0$ and 6.4 (1.0-2.5 MK), whereas the DEM-weighted average temperatures ranged between $\log T \, [K] = 6.34$ and 6.76 (2.2-5.7 MK). For all but one of the footpoint plasmoids, we find that the DEM-weighted temperature is higher than the temperature at maximum DEM (see Fig. 5, panels (a) and (b)).

- The temperature at which the DEM of the spire plasmoids peaks ranges between $\log T \, [K] = 6.0$ and 6.35 (12.24 MK), whereas the DEM-weighted average temperatures range between $\log T \, [K] = 6.14$ and 6.47 (1.33.0 MK) (see Fig. 5, panels (c)).

- A lower limit to the electron number densities of SPs (FPs) were obtained that ranged between $3.4\text{-}6.1 \times 10^8$ ($3.3\text{-}5.9 \times 10^8$) cm$^{-3}$ whereas for the spire, it ranged from $2.6\text{-}3.2 \times 10^8$ (see Table 2).

- The temperatures, DEMs (see Fig. 8) and lower limit to the electron number densities (see Table 3) were also obtained for the plasmoid as it travels along the spire (see Fig. 7). It shows the presence of low (0.5 MK) as well as high temperatures (2.5 MK). The densities as well as the peak temperatures show a systematic rise and fall during its propagation. The spire plasmoid densities range between 2.3 and $5.0 \times 10^8$ cm$^{-3}$. This explains the increase in brightness of the plasmoids over the jet spire (see Fig. 9, panel (b)).

We summarized the results obtained in Table 2 and 3 in Fig. 6 and Fig. 9 (c–f), respectively. The histograms are obtained for the peak DEM, the lower limit to electron number densities $N_e$, peak temperatures, and DEM weighted average temperatures of spire plasmoids, footpoint plasmoids, spire, and for the spire plasmoid which was tracked in Fig. 7.

Another study of plasmoids in AR jets reported by Zhang & Ji (2014b) measured plane-of-sky velocities in the range 120-450 km s$^{-1}$. These velocities are similar to those we obtained in section 3.1. They obtained DEMs for three spire plasmoids using only the six EUV AIA channels. The temperatures were ranging between 2.2–3.3 MK, whereas the lower limit to electron number densities, $N_e$ were ranging between $1.9\text{-}3.3 \times 10^9$ cm$^{-3}$. These temperatures are close to the temperatures we obtained, but the densities are an order of magnitude smaller. This might be due to the DEMs at high temperatures were not well constrained and resulting in high EM values, hence the high densities. They interpreted that the tearing-mode instability of the current sheets during the magnetic reconnection process is responsible for the formation of plasmoids along jets.

The 2D resistive magnetohydrodynamic simulations of coronal jets by Ni et al. (2017) indicated that plasmoid instability and Kelvin–Helmholtz instability are the causes of plasmoid generation. The simulations predicted multithermal plasmoids with a maximum temperature of 8 MK, lifetimes of 120 s, diameters of 6 Mm, and velocities of 200 km s$^{-1}$. We note that such temperatures are much higher than those we have measured.

The observations studied in this paper show the recurrent emission of jets and plasmoids in active regions. Previous numerical simulations showed that recurrent jets in active regions could occur due to persistent reconnection (e.g. Archontis et al. (2010), oscillatory reconnection) or flux cancellation between neighboring magnetic flux systems with oppositely directed magnetic field lines. It is likely that a similar process occurs in the jets that we have studied. Their emission continues until the reservoir of the available flux for the interacting magnetic fields is exhausted due to reconnection/flux cancellation.

Another feature of the jets studied in this paper is their elongated and curved spire, similar to that of Tripathi (2021). Numerical experiments (e.g. Gontikakis et al. 2009) showed that this could be the result of the topology between the interacting magnetic flux systems. For instance, when an emerging loop comes into contact

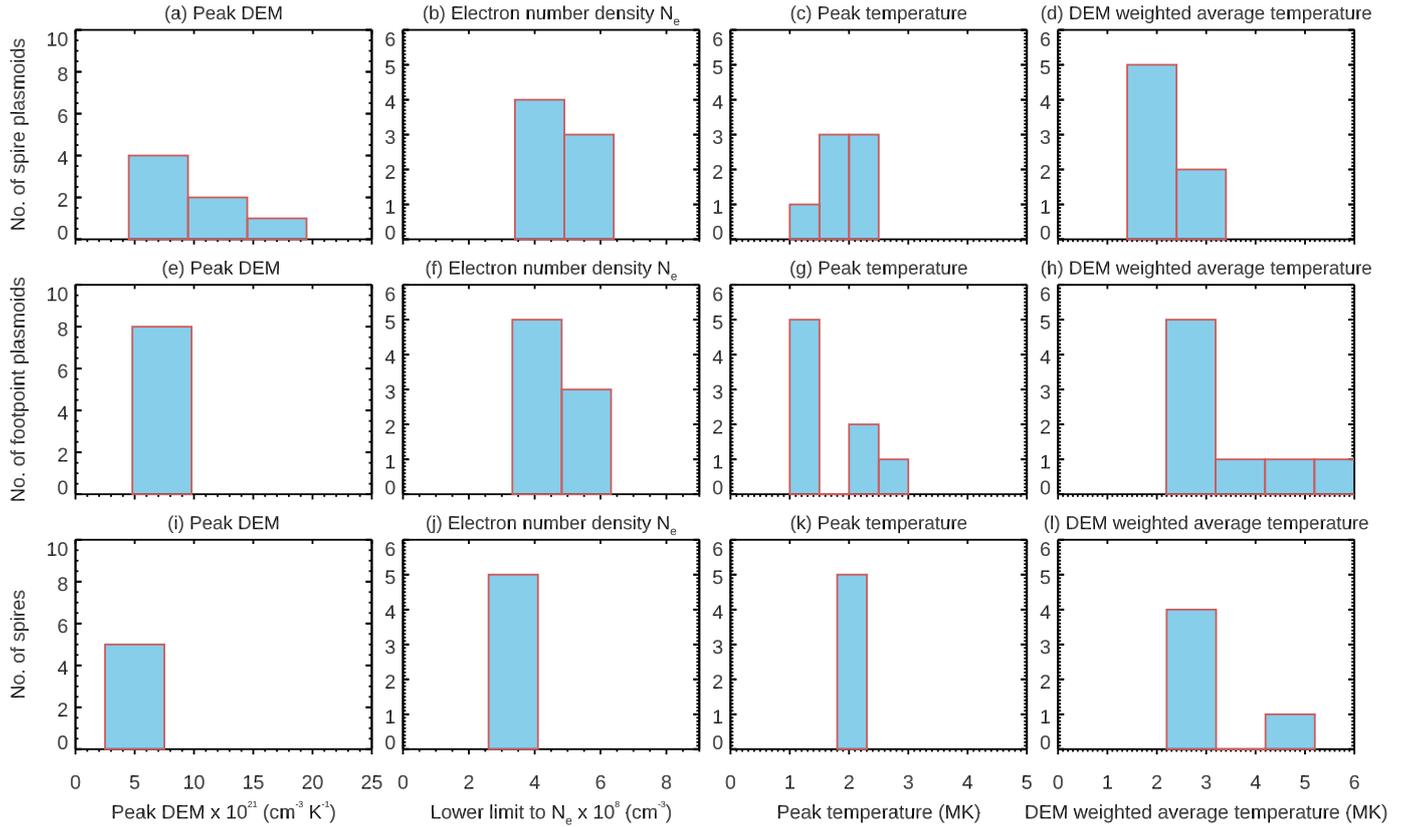

**Figure 6.** The histogram of peak DEMs (column 1), the lower limit to electron number densities, $N_e$ (column 2), peak temperatures (column 3), DEM weighted average temperatures (column 4) of spire plasmoids (row 1), footpoint plasmoids (row 2), spire (row 3).

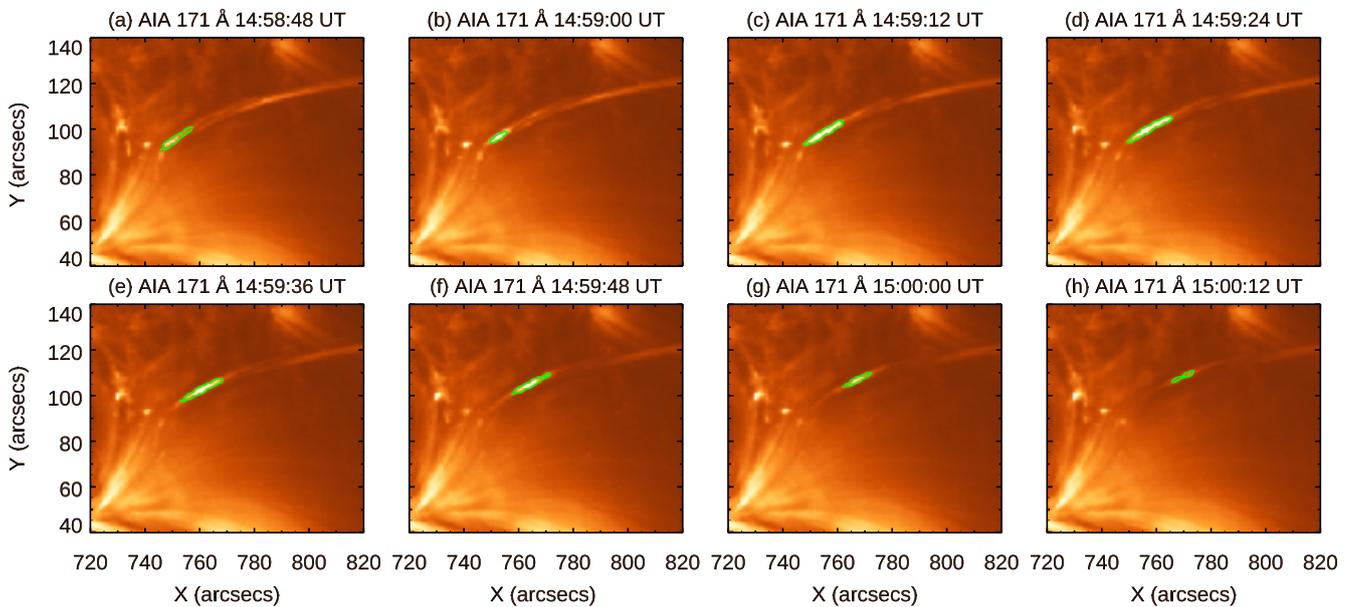

**Figure 7.** The movement of a plasmoid is tracked in successive AIA 171 Å images (re-binned AIA images to XRT resolution) during the evolution of a jet. The plasmoid along the jet spire is indicated by green contours.

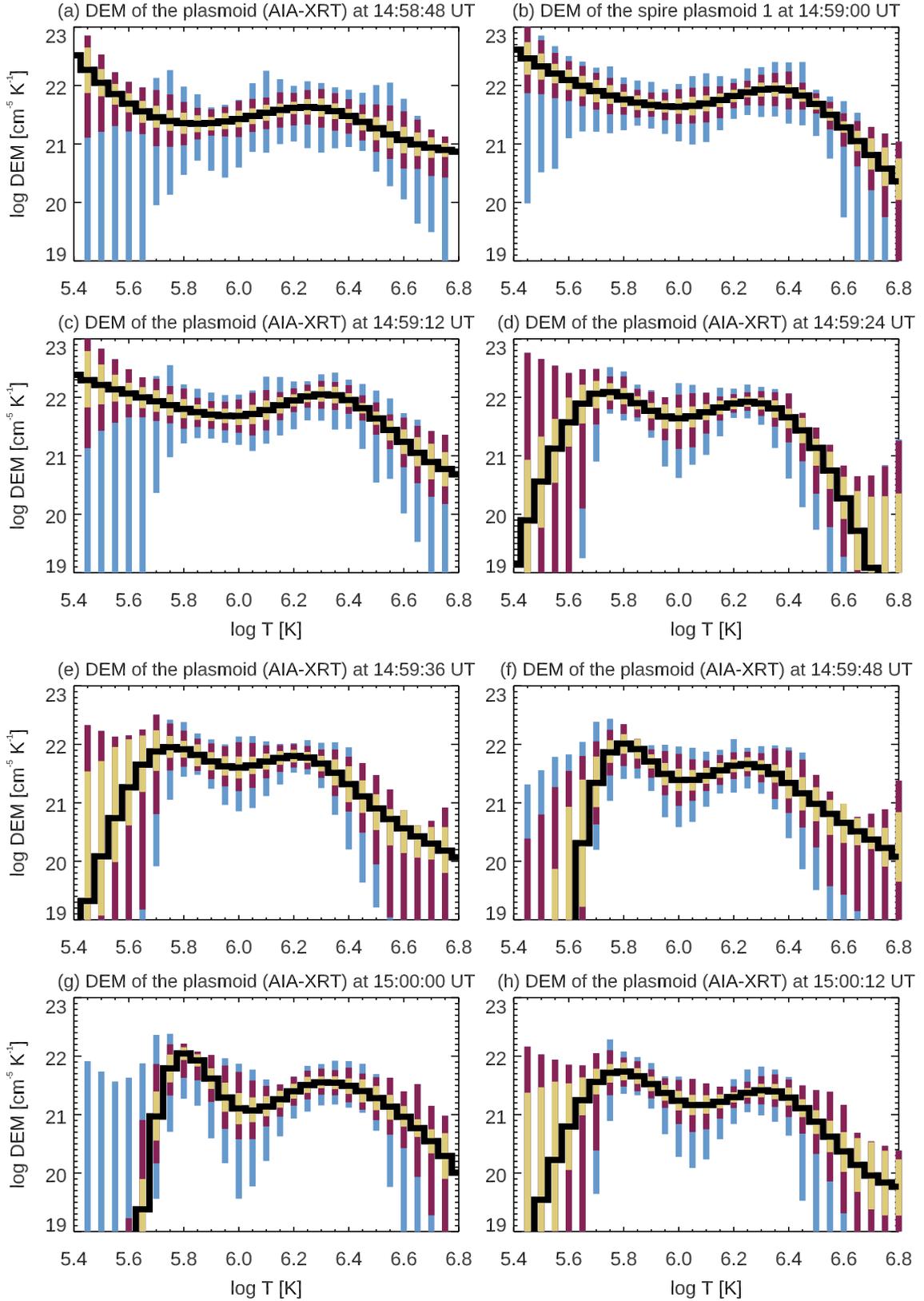

**Figure 8.** The DEM plots for the plasmoids that are shown in Fig. 7. The best-fit DEMs are shown by black curves. The yellow bars represent 50%, brown bars 80%, and blue bars 95% of the MC solutions in each temperature bin.

**Table 2.** Plasma parameters obtained for plasmoids and spire region shown in Fig. 4

| Column 1 | Column 2 | Column 3 | Column 4 | Column 5 | Column 6 | Column 7 |
|---|---|---|---|---|---|---|
| Time (UT) | plasmoids/region | peak DEM (cm$^{-3}$ K$^{-1}$) | EM at (cm$^{-3}$) | Electron number density $N_e$ (cm$^{-3}$) | log $T_{peak}$ [K] (MK) | log $\tilde{T}$ [K] (MK) |
| (a) 14:45:50 | FP1 | 8.5×10$^{21}$ | 2.5×10$^{28}$ | 4.6×10$^8$ | 6.30 (2.0) | 6.34 (2.19) |
| | SP1 | 7.2×10$^{21}$ | 4.0×10$^{28}$ | 5.8×10$^8$ | 6.20 (1.6) | 6.34 (2.2) |
| | spire | 4.1×10$^{21}$ | 1.2×10$^{28}$ | 3.2×10$^8$ | 6.34 (2.2) | 6.39 (2.5) |
| (b) 14:50:48 | SP1 | 1.0×10$^{22}$ | 4.5×10$^{28}$ | 6.1×10$^8$ | 6.20 (1.6) | 6.47 (3.0) |
| | spire | 3.1×10$^{21}$ | 8.3×10$^{27}$ | 2.6×10$^8$ | 6.35 (2.2) | 6.44 (2.8) |
| (c) 14:53:02 | FP1 | 7.9×10$^{21}$ | 1.9×10$^{28}$ | 3.94×10$^8$ | 6.0* (1) | 6.39 (2.5) |
| | FP2 | 9.7×10$^{21}$ | 2.6×10$^{28}$ | 4.6×10$^8$ | 6.4 (2.5) | 6.37 (2.3) |
| | FP3 | 9.6×10$^{21}$ | 3.4×10$^{28}$ | 5.3×10$^8$ | 6.05 (1.1) | 6.7 (5.0) |
| | SP1 | 4.5×10$^{21}$ | 1.7×10$^{28}$ | 3.8×10$^8$ | 6.0 (1.0) | 6.43 (2.7) |
| | SP2 | 1.54×10$^{22}$ | 2.6×10$^{28}$ | 4.7×10$^8$ | 6.3 (2.0) | 6.24 (1.7) |
| | SP3 | 1.3×10$^{22}$ | 1.4×10$^{28}$ | 3.5×10$^8$ | 6.2 (1.6) | 6.14 (1.4) |
| (d) 14:55:14 | FP1 | 9.7×10$^{21}$ | 4.1×10$^{28}$ | 5.9×10$^8$ | 6.1 (1.3) | 6.6 (3.4) |
| | spire | 3.9×10$^{21}$ | 1.04×10$^{28}$ | 2.9×10$^8$ | 6.35 (2.2) | 6.5 (3.0) |
| (e) 14:57:24 | FP1 | 6.0×10$^{21}$ | 3.2×10$^{28}$ | 5.2×10$^8$ | 6.1* (1.0) | 6.46 (2.9) |
| | SP1 | 5.93×10$^{21}$ | 1.37×10$^{28}$ | 3.4×10$^8$ | 6.30 (2.0) | 6.32 (2.1) |
| | spire | 2.6×10$^{21}$ | 1.2×10$^{28}$ | 3.1×10$^8$ | 6.30 (2.0) | 6.7 (5.0) |
| (f) 14:59:00 | FP1 | 5.9×10$^{21}$ | 2.5×10$^{28}$ | 4.5×10$^8$ | 6.1 (1.3) | 6.76 (5.7) |
| | FP2 | 4.8×10$^{21}$ | 1.3×10$^{28}$ | 3.32×10$^8$ | 6.30 (2.0) | 6.36 (2.3) |
| | SP1 | 7.9×10$^{21}$ | 2.9×10$^{28}$ | 4.9×10$^8$ | 6.35 (2.2) | 6.3 (2.0) |
| | spire | 2.5×10$^{21}$ | 9.3×10$^{27}$ | 2.8×10$^8$ | 6.25 (1.8) | 6.34 (2.2) |

**Notes -** FP - footpoint plasmoid, SP - spire plasmoid

**Table 3.** Plasma parameters obtained for the plasmoids shown in Fig. 7

| Column 1 | Column 2 | Column 3 | Column 4 | Column 5 | Column 6 |
|---|---|---|---|---|---|
| Time (UT) | peak DEM (cm$^{-3}$ K$^{-1}$) | EM at (cm$^{-3}$) | Electron number density $N_e$ (cm$^{-3}$) | log $T_{peak}$ [K] (MK) | log $\tilde{T}$ [K] (MK) |
| 1) 14:58:48 | 4.3×10$^{21}$ | 1.6×10$^{28}$ | 3.6×10$^8$ | 6.25 (1.8) | 6.4 (2.5) |
| 2) 14:59:00 | 8.8×10$^{21}$ | 2.83×10$^{28}$ | 4.8×10$^8$ | 6.35 (2.24) | 6.31 (2.0) |
| 3) 14:59:12 | 1.1×10$^{22}$ | 3.0×10$^{28}$ | 5.0×10$^8$ | 6.3 (2.0) | 6.4 (2.5) |
| 4) 14:59:24 | 8.4×10$^{21}$ | 1.8×10$^{28}$ | 3.8×10$^8$ | 6.25 (1.8) | 6.2 (1.6) |
| 5) 14:59:36 | 6.3×10$^{21}$ | 1.3×10$^{28}$ | 3.2×10$^8$ | 6.2 (1.6) | 6.2 (1.6) |
| 6) 14:59:48 | 4.6×10$^{21}$ | 9.4×10$^{27}$ | 2.8×10$^8$ | 6.25 (1.8) | 6.28 (1.9) |
| 7) 15:00:00 | 3.6×10$^{21}$ | 9.3×10$^{27}$ | 2.8×10$^8$ | 6.3 (2.0) | 6.36 (2.3) |
| 8) 15:00:12 | 2.6×10$^{21}$ | 7.5×10$^{27}$ | 2.3×10$^8$ | 6.3 (2.0) | 6.27 (1.9) |

with a pre-existing magnetic loop, which has already expanded into larger coronal heights, a current sheet forms at the interface between the interacting magnetic loops. Due to this topology, the interface current sheet is located at lower heights than the apex of the pre-existing magnetic loop. Then, when reconnection is triggered, the hot reconnection jet (2-3 MK) is emitted along the long and curved reconnected field lines, which are lying along the outermost layers of the pre-existing magnetic flux system. This mechanism could explain the curved spire of the observed jets and the brightening at the footpoints of the jets, which occurs due to reconnection at the lower edge of the interface current sheet (see Figs. 3 and 4 in Gontikakis et al. (2009)).

As we have discussed in the introduction, previous theoretical and numerical studies have supported the idea that the jet blobs are plasmoids, formed by an instability (e.g. tearing mode) at the interface current sheet. Our analysis shows that the emission of these plasmoids starts close to the base of the jet(s), where we believe that a strong current interface is formed as explained above. Therefore, it is likely that these plasmoids are indeed tearing-mode-induced plasmoids, which are ejected in conjunction with the hot reconnection jet. The values of the physical properties of the observed plasmoids, such as temperature, density, and velocity, are consistent with those in previous numerical experiments. For instance, it has been shown (e.g. Archontis et al. 2006) that the dynamical interaction between oppositely directed 3D magnetic fields can lead to the recurrent formation and ejection of multithermal plasmoids across the solar atmosphere (see also Isobe et al. 2007).

It is important to highlight that the physical properties of the plasmoids depend critically on the atmospheric height where they are formed. For example, the numerical experiments by Archontis et al. (2006) showed that when the plasmoids are formed at low atmospheric heights, they are usually dense (10$^7$-10$^8$ cm$^{-3}$), cool (of the order of $10^4$-$10^5$ K), and move with relatively low speeds (15-20 km s$^{-1}$). However, due to the dynamical evolution of the system, the interaction of the magnetic fields eventually occurs at larger heights too, and consequently, plasmoids are formed and emitted in the corona with higher temperatures (of the order of $10^6$ K), lower densities of 1-3×$10^6$ cm$^{-3}$ and higher velocities (150-200 km s$^{-1}$). These latter values are closer to the values we estimate based on our analysis of the observational data.

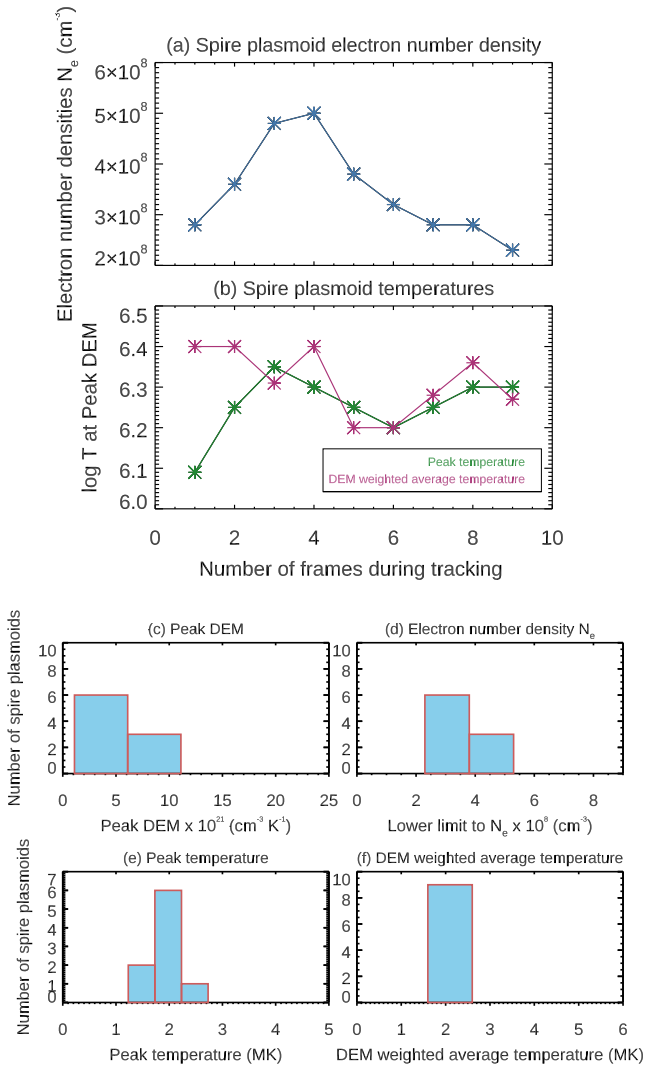

**Figure 9.** Panel (a–b): The temporal evolution of plasma parameters obtained for the spire-plasmoid, which is shown in Fig. 7. The number on the X-axis corresponds to the timings given in Table 3. Panel (c–f): The histogram of the plasma parameters given in Table 3.

The thorough investigation of the temperature and density structure of the plasmoids and spire presented here provides more accurate observational constraints for future numerical experiments.

**ORCID ID'S**

Sargam M. Mulay https://orcid.org/0000-0002-9242-2643
Durgesh Tripathi https://orcid.org/0000-0003-1689-6254
Helen Mason https://orcid.org/0000-0002-6418-7914
Giulio Del Zanna https://orcid.org/0000-0002-4125-0204
Vasilis Archontis https://orcid.org/0000-0002-6926-8676

**ACKNOWLEDGEMENTS**

Part of this work was carried out when Sargam M. Mulay (SMM) held a postdoc position at the Inter-University Centre for Astronomy and Astrophysics (IUCAA), India. Currently, SMM is a research associate at the University of Glasgow and acknowledges support from the UK Research and Innovation's Science and Technology Facilities Council under grant award numbers ST/P000533/1 and ST/T000422/1. HEM acknowledges the support of the Science and Technology Facilities Council. VA acknowledges support by the ERC Synergy Grant (GAN: 810218) 'The Whole Sun'. AIA data are courtesy of SDO (NASA) and the AIA consortium. NOAA Solar Region Summary data supplied courtesy of SolarMonitor.org. CHIANTI is a collaborative project involving George Mason University, the University of Michigan (USA), and the University of Cambridge (UK). Hinode is a Japanese mission developed and launched by ISAS/JAXA, with NAOJ as a domestic partner and NASA and STFC (UK) as international partners. It is operated by these agencies in cooperation with ESA and NSC (Norway).

**DATA AVAILABILITY**

In this paper, we used the Interactive Data Language (IDL) and SolarSoftWare (SSW; Freeland & Handy 1998) packages to analyze AIA data. Some figures within this paper were produced using IDL color-blind-friendly color tables (see Wright 2017). The AIA data is available at http://jsoc.stanford.edu/ and the data were analyzed using routines available at https://www.lmsal.com/sdodocs/doc/dcur/SDOD0060.zip/zip/entry/. The XRT data is available at http://sdc.uio.no/sdc/ and the data were analyzed using routines available at https://xrt.cfa.harvard.edu/science/tutorials.php

## APPENDIX A: MEASUREMENT OF THE WIDTH OF THE PLASMOIDS

We measured the width of the spire-plasmoids in the AIA 171 Å image at 14:53:02 UT by following the method given by Chen et al. (2022). In panel (a) of Fig. A1, we showed the active region jet in the AIA 171 channel (with pix size = 0.6") along with green contours around the spire and footpoint plasmoids. The solid blue line is plotted perpendicular to the spire-plasmoid, SP2, and the intensity profile across it is plotted in panel (b). This covers the background emission and emission from spire-plasmoid, SP2. The intensity profile was then fitted with a single Gaussian and we obtained the Full-Width Half-Maxima (FWHM). This FWHM provides us with the width of the spire-plasmoid which is 2″ i.e. 0.9 Mm (here we considered 1 AIA pixel = 750 km) which is found to be smaller than the plasmoid width measured by Chen et al. (2022).

In this observation, we noted that the width of the spire-plasmoids is the same as the width of the spire that was measured by Mulay et al. (2017a) for the same jet. The authors used spectroscopic observations from EIS/Hinode and pointed out that the emission from the spire was observed over two consecutive EIS slit positions, i.e. for 2″.

## APPENDIX B: COUNT RATES (DN/S) FOR THE PLASMOIDS

This paper has been typeset from a TeX/LaTeX file prepared by the author.

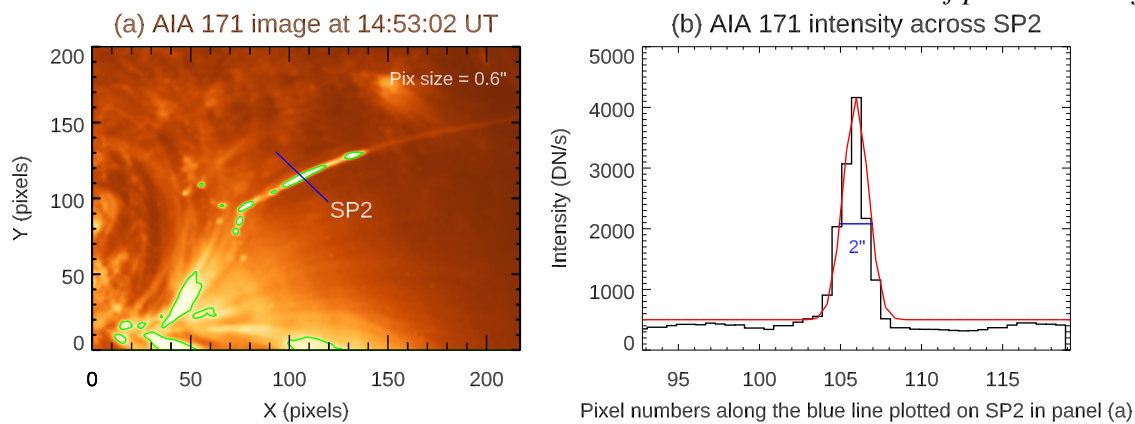

**Figure A1.** Panel (a): Active region jet at 14:53:02 UT in the AIA 171 Å channel (where 1 AIA pix = 0.6″). The plasmoids along the spire and at the footpoint regions are shown as green contours. The solid blue line is plotted perpendicular to the spire-plasmoid, SP2 and it was used to get the intensity profile across the spire-plasmoid. Panel (b): The intensity profile across the spire-plasmoid is shown as a black curve. The solid red curve is a single Gaussian fitted to the intensity profile.

**Table B1.** Observed and predicted count rates (DN/s) for the plasmoids shown in Fig. 7

| Column 1 | Column 2 | Column 3 | Column 4 | Column 5 | Column 6 | Column 7 | Column 8 | Column 9 |
|---|---|---|---|---|---|---|---|---|
| AIA/XRT | plasmoid at 14:58:48 UT | | plasmoid at 14:59:00 UT | | plasmoid at 14:59:12 UT | | plasmoid at 14:59:24 UT | |
| Channel | Observed | predicted | Observed | predicted | Observed | predicted | Observed | predicted |
| (Å) | (DN/s) | (DN/s) | (DN/s) | (DN/s) | (DN/s) | (DN/s) | (DN/s) | (DN/s) |
| AIA 94  | 15.9   | 10.4 (1.53)   | 28.5   | 12.2 (2.34)   | 23.4   | 16.0 (1.5)   | 18.1   | 7.9 (2.29)   |
| AIA 131 | 80.8   | 77.2 (1.05)   | 176.1  | 157.7 (1.12)  | 179.1  | 176.0 (1.01) | 147.9  | 163.4 (0.91) |
| AIA 171 | 1114.9 | 1198.9 (0.93) | 1948.5 | 2275.3 (0.86) | 2187.4 | 2366.8 (0.92)| 2157.1 | 2208.5 (0.98)|
| AIA 193 | 1422.9 | 1414.7 (1.01) | 2295.3 | 2505.4 (0.92) | 2562.5 | 2583.1 (0.99)| 1632.5 | 1854.9 (0.88)|
| AIA 211 | 394.7  | 545.5 (0.72)  | 800.6  | 872.7 (0.92)  | 920.5  | 981.8 (0.94) | 710    | 690.6 (1.03) |
| AIA 335 | 26.9   | 21.9 (1.23)   | 49.8   | 44.3 (1.12)   | 53.9   | 48.3 (1.1)   | 37.9   | 30.9 (1.23)  |
| Ti-poly | 171.2  | 187.3 (0.91)  | 200.2  | 233.2 (0.86)  | 263.7  | 295.3 (0.89) | 76.9   | 87.4 (0.88)  |

**Notes -** The numbers in parentheses given in Columns 3, 5, 7 and 9 indicate the ratio of observed and predicted intensities in each channel.

**Table B2.** Observed and predicted count rates (DN/s) for the plasmoids shown in Fig. 7

| Column 1 | Column 2 | Column 3 | Column 4 | Column 5 | Column 6 | Column 7 | Column 8 | Column 9 |
|---|---|---|---|---|---|---|---|---|
| AIA/XRT | plasmoid at 14:59:36 UT | | plasmoid at 14:59:48 UT | | plasmoid at 15:00:00 UT | | plasmoid at 15:00:12 UT | |
| Channel | Observed | predicted | Observed | predicted | Observed | predicted | Observed | predicted |
| (Å) | (DN/s) | (DN/s) | (DN/s) | (DN/s) | (DN/s) | (DN/s) | (DN/s) | (DN/s) |
| AIA 94  | 12.9   | 6.6 (1.95)    | 8.6    | 5.1 (1.69)    | 5.5    | 4.8 (1.5)    | 4.6    | 3.9 (1.18)   |
| AIA 131 | 110.5  | 114.9 (0.96)  | 88.05  | 62.3 (1.41)   | 79.3   | 51.7 (1.53)  | 63.4   | 90.4 (0.70)  |
| AIA 171 | 1796.6 | 1820.1 (0.99) | 1503.4 | 1326.1 (1.13) | 1361.8 | 990.1 (1.38) | 972.9  | 1019.9 (0.95)|
| AIA 193 | 1249.5 | 1414.9 (0.88) | 923.5  | 956.6 (0.97)  | 653.0  | 661.5 (0.99) | 551.1  | 628.7 (0.88) |
| AIA 211 | 498.7  | 478.9 (1.04)  | 379.2  | 359.3 (1.05)  | 353.9  | 289.8 (1.22) | 323.3  | 233.1 (1.39) |
| AIA 335 | 25.03  | 21.1 (1.19)   | 21.3   | 16.2 (1.32)   | 17.9   | 16.8 (1.07)  | 12.6   | 13.2 (0.95)  |
| Ti-poly | 63.4   | 69.2 (0.92)   | 61.6   | 67.4 (0.91)   | 91.4   | 91.3 (1.00)  | 50.3   | 52.1 (0.97)  |

**Notes -** The numbers in parentheses given in Columns 3, 5, 7 and 9 indicate the ratio of observed and predicted intensities in each channel.